\documentclass{WileyMSP-template}
\usepackage{hyperref} 
\usepackage[square,numbers,sort&compress]{natbib} 
\usepackage{float}
\usepackage{graphicx}
\usepackage{caption}
\usepackage{amsmath}

\begin{document}

\pagestyle{fancy}
\rhead{\includegraphics[width=2.5cm]{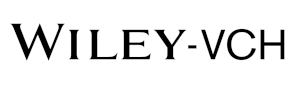}}

\title{Multimodal learning enables instant ionizing radiation alerts on unmodified mobile phones for real-world emergency response}

\maketitle


\author{Yanfeng Xie*}
\author{Xingzhi Cheng}


\dedication{}

\begin{affiliations}
Yanfeng Xie\\
Radiation Sciences Graduate Program, McMaster University, Hamilton, ON, L8S 4K1, Canada\\
Email Address: xie22@mcmaster.ca

Dr. Xingzhi Cheng\\
Radiation Sciences Graduate Program, McMaster University, Hamilton, ON, L8S 4K1, Canada\\

\end{affiliations}



\begin{abstract}

In a radiation emergency, every second counts, yet the public rarely has immediate access to dedicated monitoring devices when they are needed most. Here, the first practical mobile phone-based emergency ionizing radiation detection method is presented that operates entirely without requiring camera coverage or additional hardware modifications. Utilizing a multimodal deep learning approach that integrates sparse radiation-induced signals distributions with the brightness patterns, the proposed framework effectively isolates subtle radiation signals from overwhelming visual interference. A hybrid 3D-2D convolutional neural network (CNN) identifies radiation-induced spots from raw mobile phone video, while a multi-layer perceptron (MLP) fuses the radiation signal and brightness maps for the dose rate estimation. The method detects hazardous dose rates (25$\sim$280 mRem/h) rapidly within six seconds (accuracy 86$\sim$96\%), and low-level radiation ($\sim$0.6 mRem/h) with extended measurement durations achieves 87\% accuracy. The developed method greatly enhances mobile phone radiation detection practicality and shows substantial potential as an accessible radiation emergency detection tool.

\end{abstract}


\section{Introduction}


All individuals are continuously exposed to background radiation originating from cosmic and solar radiation, as well as naturally occurring radionuclides in the Earth's crust. The health risks associated with such natural background radiation are minimal. Moreover, regulatory frameworks have established dose limits to ensure public safety from artificial sources of radiation. In Canada, legislation mandates that the annual effective dose to members of the general public must not exceed 1 mSv \cite{CNSC_RadiationHealthEffects, Canada_SOR2000_203}.\\

In situations involving uncontrolled or previously unidentified sources of radiation, especially prior to official recognition by authorities, the general public often lacks effective means of protection against radiological hazards. A notorious example is the Goiânia accident in 1987, where an abandoned teletherapy source was dismantled and stolen by scavengers, resulting in the contamination of over 200 individuals and causing four death \cite{IAEA_Goiania_1988}. Between 1980 and 2013, a decline in the number of reported radiation accidents and overexposed individuals has been observed, likely attributable to sustained advancements in radiation protection practices \cite{Coeytaux2015_PLoSOne}. Nevertheless, in recent years, incidents involving the loss of radioactive sources continue to highlight ongoing risks. Notable examples include the Caesium capsule incident in Western Australia in 2023 \cite{BBC_2023_capsule}, the Prachin Buri cylindrical Caesium source loss in 2023 \cite{CNN_2023_ThailandCylinder}, and the Khabarovsk radioactive capsule incident in 2024 \cite{NDTV_2024_Khabarovsk}. Accurate detection of ionizing radiation contamination or hazards typically requires the use of specialized instrumentation such as Geiger-Müller counters, ion chamber survey meters, or electronic personal dosimeters. In emergency situations, such instruments are typically not immediately available to the general public.\\

Therefore, in response to uncontrolled radiation emergencies, mobile phone-based radiation detection algorithms have been developed to address the gap in immediate public access to monitoring tools, given that most individuals consistently carry mobile phones equipped with CMOS image sensors capable of responding to both visible light and ionizing radiation. The earliest implementation of radiation detection using mobile phones was proposed by Matthias et al. in 2012 \cite{Michelsburg2012_SPIE}. Since then, the concept has attracted widespread attention, with multiple studies evaluating the feasibility and performance of mobile phone radiation detection systems \cite{Johary2021_SciRep, Huang2020_OpenPhysics, Ishigaki2013_IEEEJSEN}. These studies typically require the phone camera to be covered with opaque material, such as black tape or other physical shielding. Although some research suggests that camera covering may not be essential, those methodologies still necessitate specific positioning of the device—such as placing it on a surface with the rear camera facing downward or positioning it in a pocket—to effectively suppress visible light interference and isolate radiation-induced signal patterns \cite{Wei2017_NST}. Such operational constraints significantly reduce the practical applicability of mobile phone-based radiation detection, as controlled camera covering is often impractical in real emergency situations. Furthermore, none of the applications described remain functional or publicly available, likely due to technological changes over the past decade. In particular, the progressive miniaturization of CMOS sensors in modern mobile phones—driven by advancements in image resolution—has resulted in higher electronic noise and increased challenges in radiation signal discrimination.\\
This study targets modern mobile phone platforms. Using the iPhone 15 Pro as a representative example, a radiation dose rate prediction framework based on multimodal deep learning has been developed on PyTorch \cite{Paszke2019_PyTorch}, which operates without requiring any physical camera covering or additional hardware modifications. To construct the training database, various intervening materials were placed between the mobile phone camera and the radiation source, allowing the model to learn to suppress visible light interference and focus on radiation-induced signal patterns. This approach enables rapid identification of hazardous radiation levels exceeding 280 mRem/h within a measurement window of six seconds. Additionally, for lower-risk scenarios involving items suspected of emitting low-level radiation ($\sim$0.6 mRem/h), the model can achieve a dose prediction accuracy of up to 87\% by extending the acquisition time to 50 minutes.\\

\begin{figure}[H]
\centering
\includegraphics[width=0.8\textwidth]{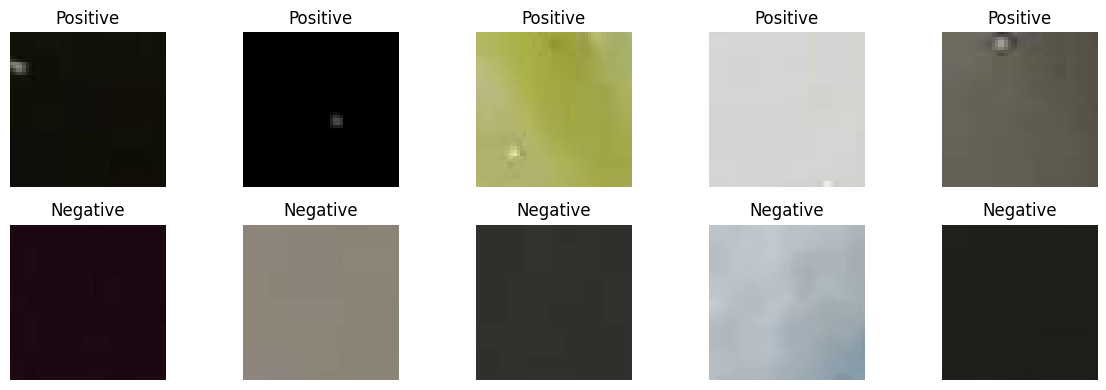}
\caption{Video segmentation frames visualization with (Positive) and without (Negative) RC spot}
\label{fig:rc_spot}
\end{figure}

The method utilizes a hybrid 3D-2D spatio-temporal convolutional neural network (the 3D CNN) to analyze cropped segments from the raw video stream. The network classifies these segments to identify the presence of speckle-like patterns caused by ionizing radiation—termed Radiation-caused Spots (RC Spots), as illustrated in Figure \ref{fig:rc_spot}. The classification results are then aggregated to form a radiation signal distribution matrix representing the entire video. This matrix is subsequently fused with the corresponding brightness matrix extracted from the original footage, and the combined information is processed using a multi-layer perceptron to predict the final dose rate. The proposed framework demonstrates consistent performance on an independent external dataset. Although further validation with expanded datasets and broader environmental testing is required prior to the public release of a robust mobile application, this approach shows substantial potential to support real-time detection of radiation hazards. It may assist users in avoiding exposure during emergency situations and help ensure that any received dose remains within the regulatory annual limit for public exposure, thereby enhancing safety and situational awareness.

\section{Results \& Discussion}

\subsection{Principal Verification and Physics simulations}
\begin{figure}[H]
\centering
\includegraphics[width=1\textwidth]{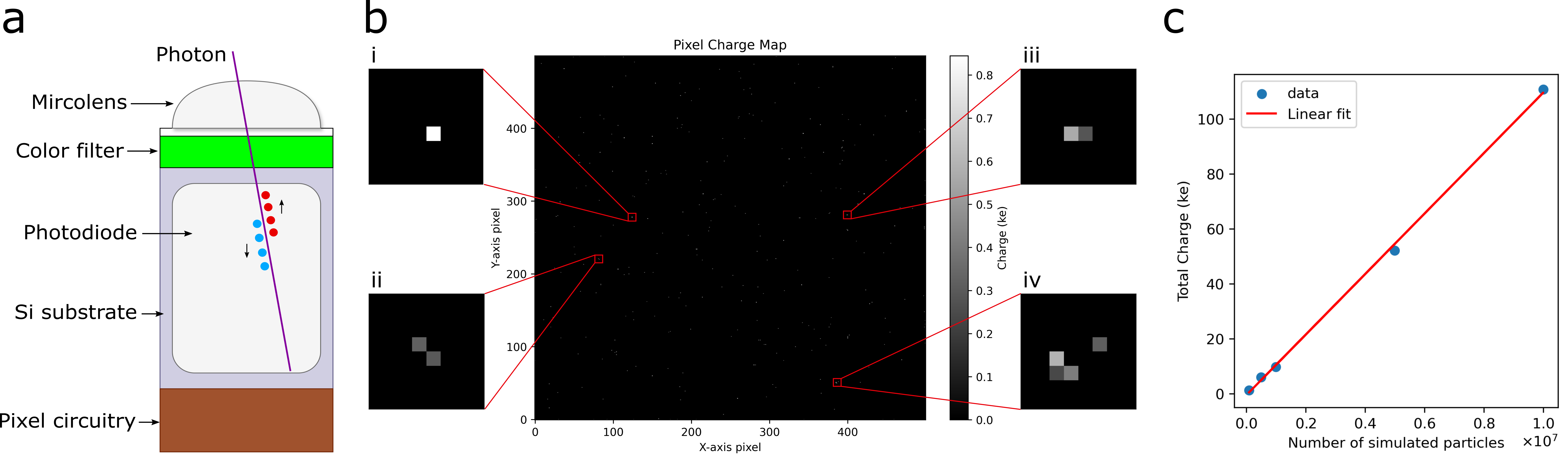}
\caption{(a) Illustration of a CMOS imaging sensor pixel. Red and blue spheres represent the electrons and holes, respectively. (b) Pixel charge map depicting the spatial distribution of charge deposition across a CMOS imaging sensor. Subfigures i–iv provide magnified views of regions highlighted by red squares, showing localized charge distribution. The X and Y axes denote pixel coordinates in the horizontal and vertical directions, respectively. Grayscale intensity, as indicated by the color bar, quantifies the collected charge in kilo-electrons (ke). (c) Linear relationship between the number of simulated particles and the total collected charge from the CMOS pixel sensor.}
\label{fig:cmos}
\end{figure}

The complementary metal-oxide semiconductor (CMOS) image sensors are widely used for mobile phones nowadays. The general structure of a CMOS sensor pixel is shown in Figure \ref{fig:cmos}a.  After photons entering the microlens, they pass through the color filter and reach the photodiode, a reversely biased silicon diode. The energy deposited by photons within the photodiode’s depletion region leads to the production of electron-hole pairs which drift along the electrical field and are collected by electrodes. While visible light photons are fully and locally absorbed via the photoelectric effect within the photodiode's active volume, energetic gamma rays are rarely absorbed within a single photodiode cell due to the extremely small active volume. Instead, the charge carriers are typically formed by energy deposition of photoelectrons, Compton electrons, and other secondary electrons originating from gamma interactions in the CMOS pixel or auxiliary structures. The accumulated charge stored in the transistors is then transferred as voltage signal to the analogy-to-digital converter (ADC) which gives the pixel values for imaging formation. In the case of radiation source presents, bright spots or blotches are found in the photo or a frame of the video. \\

Allpix\textsuperscript{2} simulations were carried out to further understand the radiation effects on a CMOS imaging sensor. Allpix\textsuperscript{2} is an open-source, easy-to-use simulation framework developed at CERN for semiconductor pixel sensors utilizing Geant4 and ROOT toolkits \cite{Allpix2_2018}. \\

A charge map that describes the distribution of charge accumulation in pixels for $10^7$ source photons was plotted as shown in Figure \ref{fig:cmos}b. The map predominantly shows a black background, indicating pixels with no charge deposition or below the digitization threshold. Superimposed are isolated bright (white to gray) blotches. These discrete spots correspond to individual fired pixels or small clusters where are predominantly attributed to localized energy deposition events. At gamma energy of 59.5\,keV, the photoelectric effect is the dominant interaction mechanism, leading to the creation of electron-hole pairs concentrated within a single pixel (Figure \ref{fig:cmos}b (i)) or a very confined pixel cluster (Figure \ref{fig:cmos}b (ii) and (iii)). Less frequent Compton scattering events could also contribute, potentially involving slight charge sharing over a few adjacent pixels together with the IPC effect, for example, the pattern shown in Figure \ref{fig:cmos}b (iv).

Because the source geometry is a circular beam inscribed in the sensor active area, most of the blotches are found within the beam radius. However, there are a few blotches that can be observed at the conners of the sensor which might be produced by scattered gamma rays. With $10^7$ source particles, 311 hits and $110\,\mathrm{ke}$ total charge deposition are registered. Despite the low detection efficiency, the CMOS pixel sensor provides a direct means for radiation monitoring.

To investigate the relationship between the response of the sensor to the radiation field strength, a series of simulations were conducted for the same configuration described above but with different source particles numbers ranging from $10^5$ to $10^7$. Figure \ref{fig:cmos}c demonstrates a highly linear relationship between the number of simulated particles and the total charge deposition in the sensor, which is also proportional to the number of bright spots. The linear proportionality between the radiation source strength and the number of blotches from the image lays the foundation of using mobile phone camera CMOS sensor as a radiation dosimeter.

\subsection{Multimodal Learning Workflow}

Due to the intrinsically low interaction cross-section between ionizing radiation, particularly gamma photons, and the CMOS image sensors employed in mobile phone cameras, the radiation-induced signals, referred to as RC spots, are generally extremely sparse. Even with a 555 MBq radioactive source, only a limited number (10 per second) of RC spots can be captured. Extracting such rare signals directly from raw video data using machine learning poses a significant challenge, especially given the overwhelming volume of visual information present without any camera covering. 

To mitigate this difficulty, the video data is partitioned into smaller segments along both spatial and temporal dimensions, allowing each segment to be individually analyzed. This segmentation strategy is justified by the fact that RC spot occurrence is independent of the visible light content in the captured frames and, in addition, occurs randomly in space, resulting in a theoretically uniform spatial distribution. Furthermore, it enables substantial expansion of the training dataset for machine learning purposes. 

Building upon this concept, a multimodal deep learning workflow has been developed for comprehensive mobile phone-based radiation detection, as illustrated in Figure~\ref{fig:workflow}. The methodology comprises five stages:
(a) Segmentation, where the video is divided into smaller temporal-spatial units; (b) Data Preprocessing, which includes normalization and data reshaping; (c) 3D CNN Scanning Analysis, whereby each segment is assessed using the 3D CNN to identify RC spot patterns; (d) Brightness and Signal Distribution Map Generation, producing structured representations of radiation signals and ambient lighting conditions; and (e) Final Dose Rate Prediction, where the fused information is analyzed using a multi-layer perceptron to estimate the dose rate. 

This modular architecture provides a robust framework for sparse radiation signal identification under practical conditions and contributes to the advancement of emergency radiation sensing using standard mobile phone hardware.

\begin{enumerate}
\item[(a)] \textbf{Segmentation:} The segmentation strategy is guided by empirical upper bounds on RC spot occurrence within both spatial and temporal domains during data acquisition. Accordingly, videos recorded by mobile phones are divided into multiple short clips, each comprising a $50\times50$-pixel region, with three color channels and eight frames.

\item[(b)] \textbf{Data Preprocess:} Each segmented clip is processed using a Channel-Time Folding approach, where color channels and temporal frames are rearranged and flattened into a unified three-dimensional matrix of size $50\times50\times24$. This transformation not only improves disk input/output efficiency but also ensures compatibility with the PyTorch framework.

\begin{figure}[H]
\centering
\includegraphics[width=0.8\textwidth]{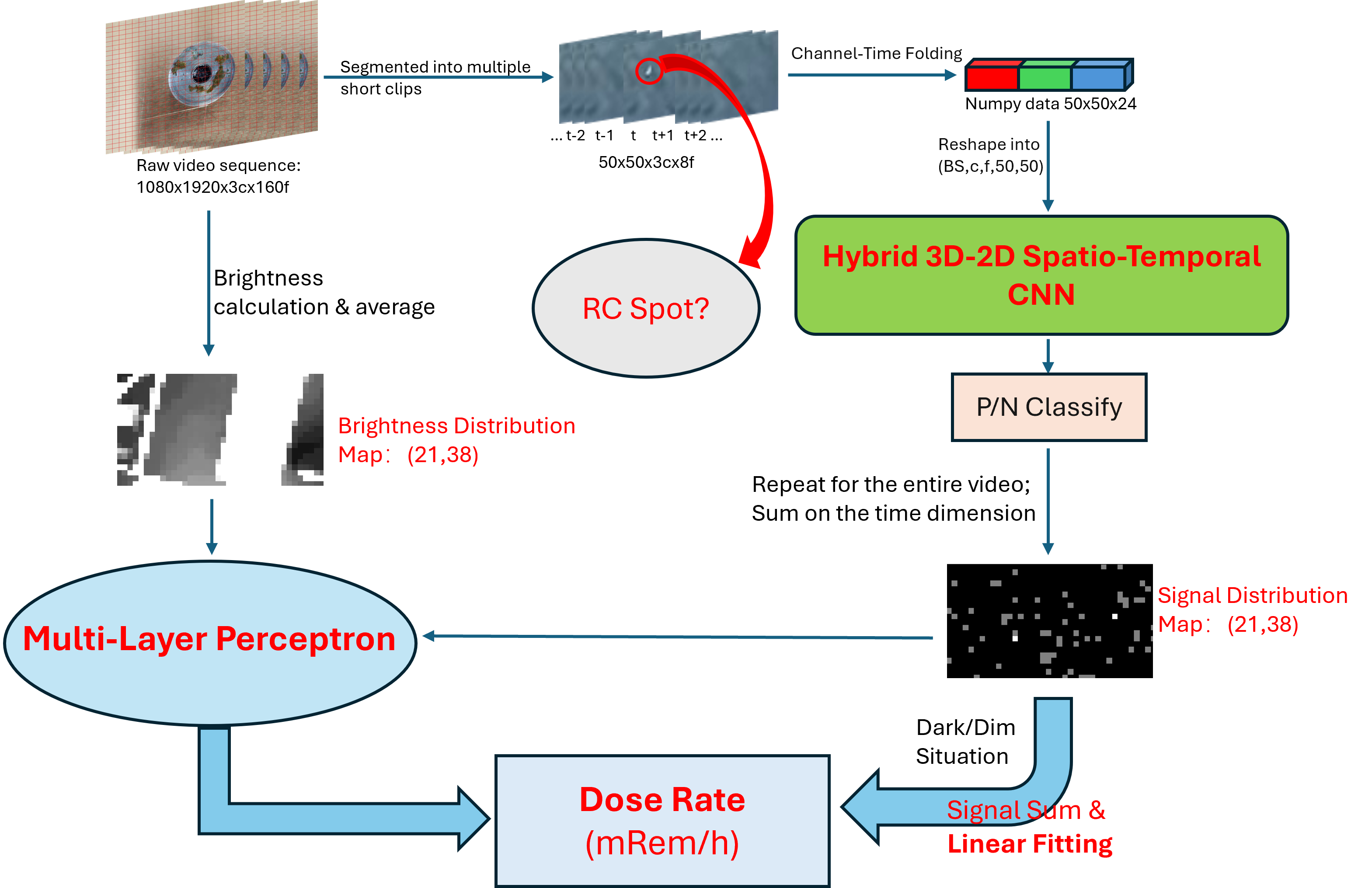}
\caption{Multimodal Learning Workflow for mobile phone radiation detection: Segmentation, Data Preprocess, 3D CNN Scanning Analysis, Brightness \& Signal Distribution Maps Obtain, Final Dose Rate Prediction.}
\label{fig:workflow}
\end{figure}

\item[(c)] \textbf{3D CNN Scanning Analysis:} The processed clips are subsequently passed into a lightweight hybrid 3D-2D spatio-temporal convolutional neural network (3D CNN) for binary classification. Each video segment is classified as either Positive (containing an RC spot) or Negative (no RC spot), with further architectural and training details to be presented in the following subsection.

\item[(d)] \textbf{Brightness \& Signal Distribution Maps Obtain:} The trained 3D CNN model is applied across the entire raw video, which has a dimension of $1080\times1920\times3$ (channels) $\times160$ (frames). This analysis yields a two-dimensional RC spot signal distribution map of size $21\times38$, where the dimensions correspond to the number of non-overlapping $50\times50$-pixel regions that fit within the $1080\times1920$ frame ($\lfloor1080/50\rfloor = 21$, $\lfloor1920/50\rfloor = 38$). To achieve uniform segmentation, small boundary regions are cropped, resulting in a negligible loss of peripheral information. The values in the map are obtained by aggregating spatial detections and summing over the temporal axis. In scenarios involving weak lighting conditions or objects with simple surface patterns---such as uniformly colored sticky notes---the signal distribution map is often unnecessary, as the 3D CNN demonstrates high classification accuracy, with a false positive rate (FPR) as low as 0.03\% and a recall rate up to 97\%. In such cases, the radiation dose rate can be directly estimated using the linear relationship between dose rate and total RC spot count. Nevertheless, in more general and complex detection scenarios, the performance of the 3D CNN may become unstable due to visually induced signal-like artifacts in the input video. For instance, flashing animations or glossy materials under strong illumination may introduce flickering regions that are incorrectly identified as RC spots. Such false signals exhibit spatial clustering, as they stem from visual features rather than random radiation interactions.

\begin{figure}[H]
\centering
\includegraphics[width=0.8\textwidth]{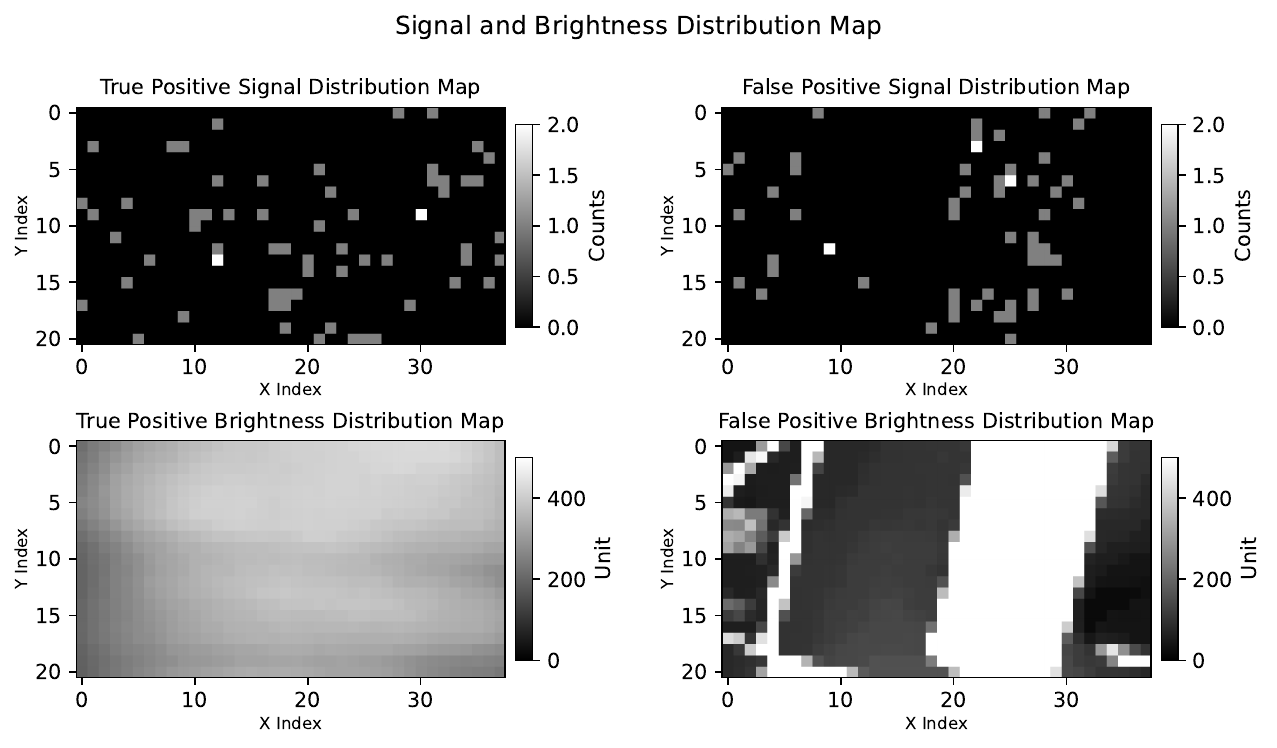}
\caption{The compassion between a True Positive Signal/Brightness Distribution Map and a False Positive Signal/Brightness Distribution Map with similar total signal counts (65 vs 58).}
\label{fig:map}
\end{figure}

In contrast, true RC spots are distributed more uniformly across the sensor due to the stochastic nature of radiation interactions with the CMOS layer. Figure \ref{fig:map} compares a true positive signal distribution map with a false positive map, both containing comparable total signal counts (65 vs.\ 58). In the false case, signals are predominantly concentrated between X-indices 20 to 30, which aligns closely with regions of increased brightness in the corresponding brightness distribution map. This correlation can be effectively quantified using statistical analysis. Consequently, the brightness distribution map, calculated by averaging pixel intensity across both spatial and temporal dimensions, is introduced as an additional input feature to suppress the impact of false positives on dose estimation.

\item[(e)] \textbf{Final Dose Rate Prediction:} A dual-branch multi-layer perceptron (MLP) is developed to perform the final dose rate estimation. The first branch processes the flattened two-dimensional signal and brightness distribution maps ($21\times38\times2$), while the second branch ingests seven engineered statistical features, including total signal count, mean and standard deviation of both signal and brightness, the signal--brightness correlation coefficient, and the signal uniformity metric. The signal--brightness correlation coefficient and the signal uniformity metric are both defined in subsection 4.3 Quantification and Calculation Methods. These features are designed to explicitly capture the relationship patterns highlighted above, particularly the suppression of visually induced false positives by quantifying their correlation with local brightness patterns.

Each branch comprises two fully connected layers with ReLU activation, accompanied by moderate $L_2$ weight regularization and dropout rates between 0.25 and 0.35 to prevent overfitting while preserving learning capacity for the relatively small dataset ($\sim$3000 samples). The outputs of the two branches are concatenated and passed through two additional dense layers before a final linear neuron outputs the predicted dose rate. The performance of this prediction framework is evaluated in detail in subsection 4.4 Dose Rate Prediction Evaluation.
\end{enumerate}

\subsection{Hybrid 3D-2D Spatio-Temporal CNN}

\begin{figure}[H]
\centering
\includegraphics[width=0.9\textwidth]{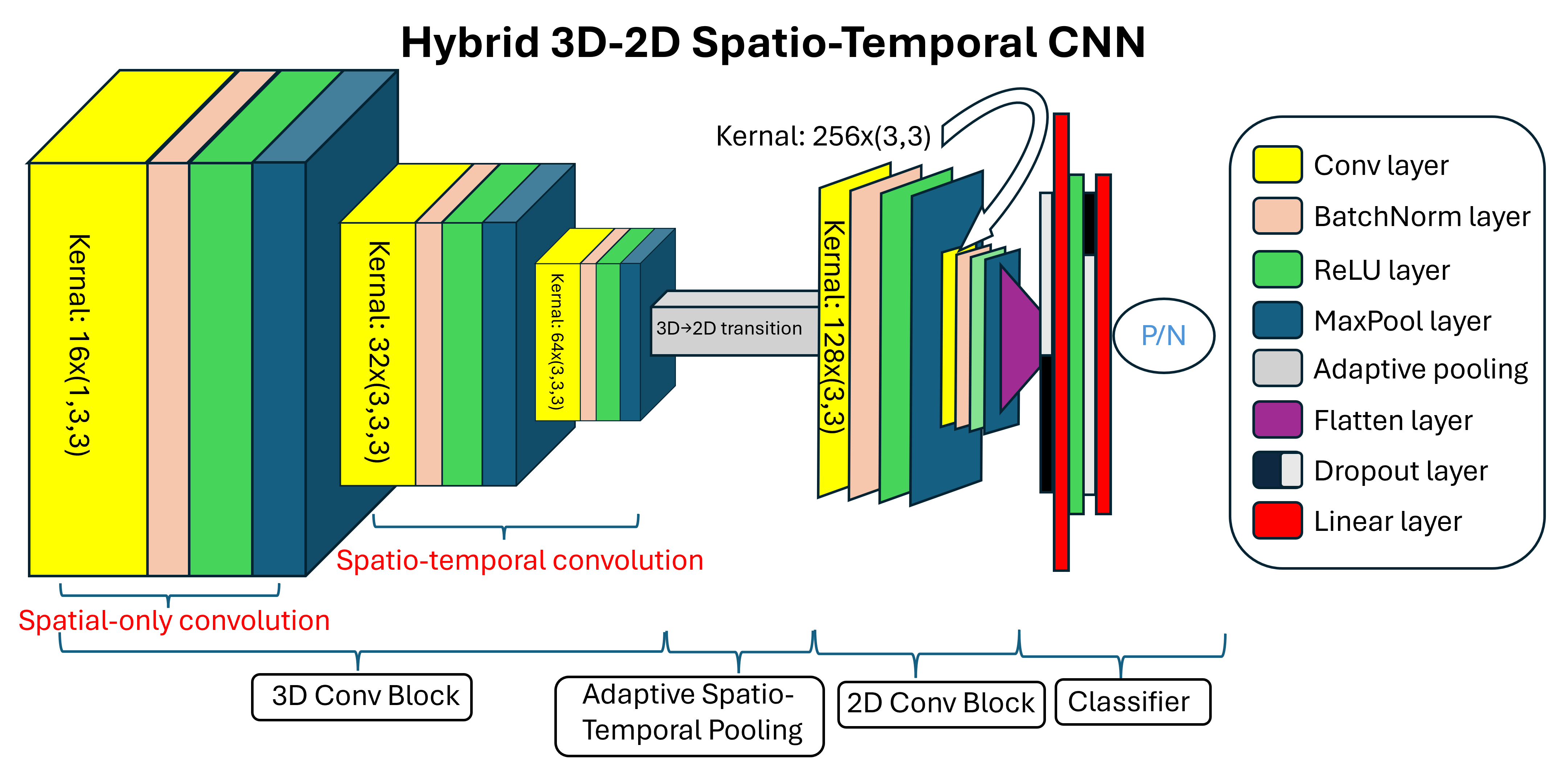}
\caption{Hybrid 3D-2D Spatio-Temporal CNN structure illustration: 3D Conv Block, Adaptive Spatio-Temporal Pooling, 2D Conv Block, and the final Classifier.}
\label{fig:cnn}
\end{figure}

To accurately classify whether a cropped video segment contains an RC spot, the classification algorithm must be capable of extracting both spatial and temporal features, as RC spots typically appear in a single frame and disappear in the next. In conventional image-processing pipelines, transient point-like anomalies in relatively static videos may be detected using background modeling and subtraction methods (e.g., Gaussian mixture models \cite{Stauffer1999_CVPR} or frame differencing \cite{Collins2000_CMURI}). While these approaches are effective for detecting clear, high-contrast changes, they are less suited for RC spots—transient, single-frame signals induced by ionizing radiation—whose extremely low intensity makes them difficult to distinguish from fine-scale surface patterns inherently present in the scene under visible light. The challenge is further compounded in non-static videos, where object surfaces may exhibit complex textures (e.g., blotches on contaminated wiping paper similar in size to RC spots) and geometric motion, both of which can generate background variations that interfere with reliable detection. The proposed hybrid 3D-2D CNN is trained to learn discriminative spatiotemporal features and is able to detect subtle and transient events without the need for manually crafted background models. Its architecture is illustrated in Figure.\ref{fig:cnn}.

The model operates in three distinct phases. In Phase One, three stacked 3D convolutional layers are applied, each followed by Batch Normalization and ReLU activation. The first convolutional layer focuses exclusively on spatial feature extraction, while the second and third layers implement spatio-temporal convolutions to jointly process spatial and temporal dynamics. Each convolutional layer is followed by max-pooling, resulting in a coarse-to-fine spatio-temporal feature hierarchy and a progressive reduction in input dimensions (from $50\times50\times8$ to $12\times12\times4$). Phase Two begins with an adaptive average pooling operation that standardizes the output to a fixed resolution of $1\times4\times4$. The temporal dimension is then squeezed to prepare the data for the final processing phase. In Phase Three, 2D convolutional layers are used to refine abstracted spatial features, ultimately reducing the output to a global spatial resolution of $1\times1$ for final binary classification. This architecture balances temporal modeling capability with computational efficiency.

To prevent the 3D CNN from learning object-specific visible light features---such as surface patterns or distinguishable shapes---various surface-blocking materials were introduced between the mobile phone and the radiation source during data collection for database construction. These materials including, but are not limited to, memo paper, printed cards, cardboard boxes, plastic bags, disposable face masks, and toilet paper, are used for both positive and negative database building. The resulting dataset comprises 8,824 positive and 20,000 negative samples. The theoretical lower limit of detectable dose rate is directly related to the classifier's FPR and recall; for practical use to identify radioactive source of 60 mRem/h, a classifier should achieve at least FPR $< 0.3\%$ and recall $> 60\%$, which is further discussed in the following subsection. The trained 3D CNN demonstrates strong performance on the reserved test dataset, achieving a false positive rate below $0.2\%$ and recall above $99.3\%$. In parallel, a Vision Transformer (ViT) model was also trained. Unlike CNNs, ViT applies self-attention mechanisms that do not impose local inductive biases and may, therefore, focus dynamically on global anomalies such as RC spots across the entire video frame. However, the trained ViT model achieved only moderate performance, with a FPR of approximately $10\%$ and recall of about $90\%$. These results align with existing findings that CNNs typically outperform transformers when applied to relatively small-scale datasets \cite{Lee2021_ViT_SmallDatasets, Liu2021_EfficientViT_SmallData}. This suggests that the current database remains insufficient for fully leveraging transformer architectures and that model performance could be further enhanced by expanding the dataset in the future work.\\

To further evaluate the generalization capability of the 3D CNN, an independent external validation dataset was collected using a distinct and previously unseen surface-blocking material---corn husk---under varying brightness conditions. The model achieved excellent results on this dataset, with an FPR below 0.03\% and recall above 97\%. Nevertheless, the generalization power of this evaluation is limited due to the small sample size: 20,000 negative and only 94 positive samples, constrained by the high cost of database generation. Given the virtually infinite variability in object surface patterns and ambient lighting conditions, it is possible that specific scenarios may trigger false positive signals not captured by this limited validation set. Therefore, to enhance robustness and mitigate such risks, integrating brightness information through the proposed multimodal learning workflow remains essential. This strategy provides an effective approach to account for potential false positive and false negative cases, supporting more reliable dose estimation in real-world applications.

\subsection{Dose Rate Prediction Evaluation}

\threesubsection{General prediction evaluation}

The model demonstrates reliable performance on the test dataset when either the predicted or the actual dose rate exceeds the threshold of 25 mrem/h. Under these conditions, the model provides accurate numerical estimates, achieving an average prediction accuracy of 86.7\% and the lowest observed performance of 50\%. Moreover, it correctly classifies all tested video samples (over 200 cases involving various dose levels and object surfaces) as either above or below the 25 mrem/h threshold. This threshold is constrained by the physical sensitivity limits of the detection hardware, the measurement duration and the 3D CNN performance. Here, prediction accuracy is expressed as one minus the relative deviation between the predicted and true values.\\

This level of accuracy is sufficient for rapid radiation hazard identification and practical use in public emergency scenarios. In practical use, if the estimated dose rate near a suspicious object exceeds 25 mrem/h, a prudent individual precaution is to promptly increase distance from the object or location. This is because the annual public dose limit is 100 mrem; at a dose rate of 25 mrem/h, the limit would be reached in just four hours of continuous exposure. It is important to note that in realistic emergency situations, individuals may only recall the availability of a radiation detection app after several minutes or even hours of exposure. Furthermore, in the rare event of a nuclear incident, radioactive contamination is likely to be widespread near the source, while a mobile phone camera can only detect radiation particles that happen to strike its relatively small sensor area. As a result, the actual dose rate experienced by the user may be underestimated. Given these constraints, the proposed multimodal learning method should be regarded primarily as a hazardous radiation source alert tool rather than a precise dosimeter. While this method can provide reasonably accurate estimates of higher dose rates, the central safety message for public use remains the same: once the estimated dose rate exceeds 25 mrem/h with high confidence, individuals should take timely personal protective action (such as moving away from the source) to minimize the risk of exceeding the annual public dose limit.\\

The accuracy of dose rate prediction is influenced by several limiting factors, including mobile phone hardware specifications and the benchmark instrument used during database construction. The ion chamber survey meter employed in the data collection process exhibits a measurement uncertainty of approximately 30\%, primarily due to unavoidable small displacements when positioning the device for dose rate measurements. Another significant constraint is the limited and discontinuous range of dose rate levels used during mobile phone data acquisition, which affects the granularity of the training dataset. Additional experimental details are discussed in Section 4.\\

\threesubsection{Comparison between MLP and Linear fitting}

In the absence of the multimodal learning workflow, the theoretical dose rate detection threshold can be estimated based on the standalone performance of the 3D CNN. With a FPR of approximately 0.2\% for general object surface patterns and ambient brightness conditions and given that a single 160-frame video clip generates 15,960 cropped segments, approximately 30 false signal detections are expected per video. To achieve a 90\% one-sided confidence level in determining whether hazardous radiation is present (i.e., exceeding 25 mrem/h), the critical level analysis indicates that more than 40 RC spot counts are required. The critical level calculation detail is included in subsection 4.4. Empirical results suggest that 40 true signal counts correspond to a dose rate of approximately 55 mRem/h, which is much worse than the model’s actual detection limit achieved through multimodal learning.

\begin{figure}[H]
\centering
\includegraphics[width=0.9\textwidth]{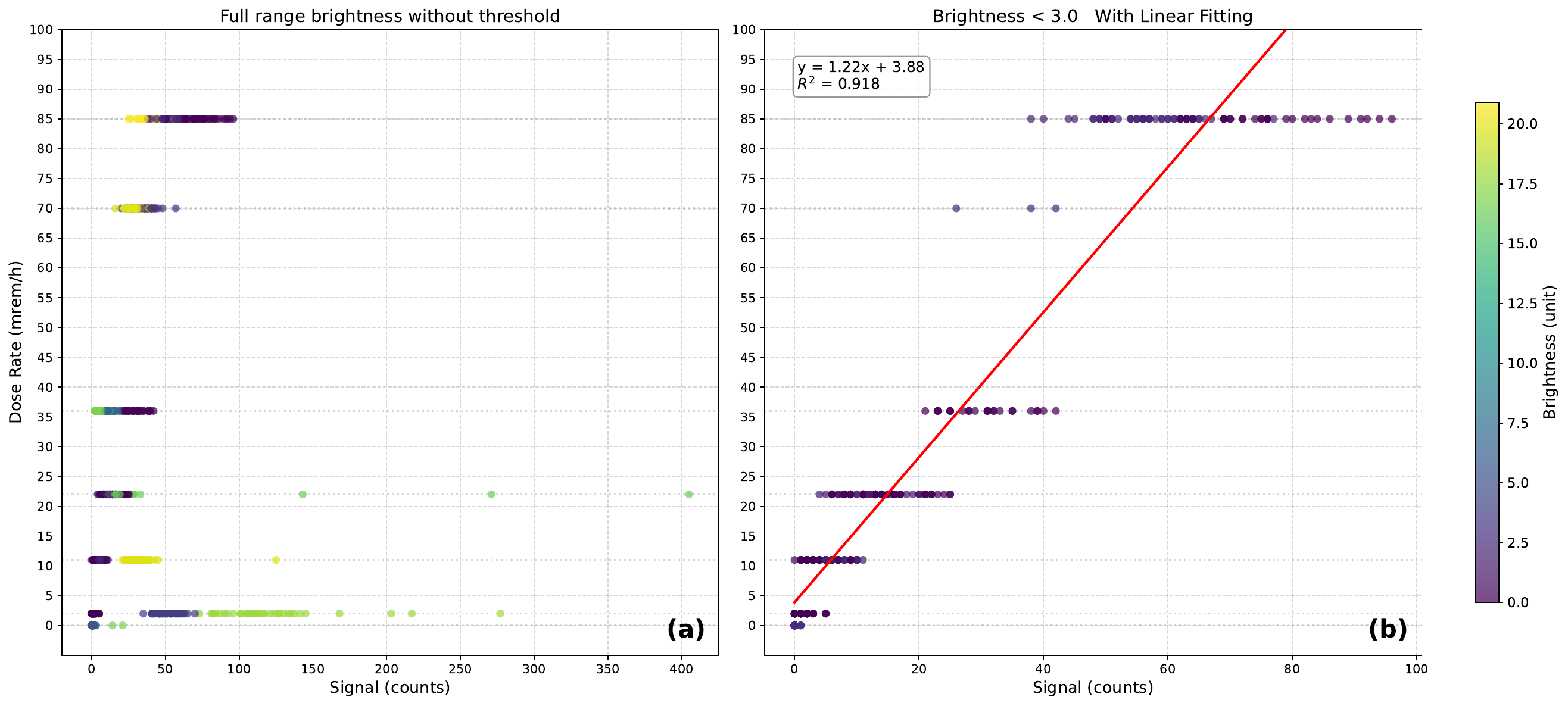}
\caption{Relationship between dose rate and signal counts with and without brightness thresholds.}
\label{fig:linear}
\end{figure}

This improvement is largely attributable to the multi-layer perceptron (MLP) component, which incorporates brightness distribution information to suppress false signal contributions. As illustrated in Figure.\ref{fig:linear}a, under typical lighting and complex surface conditions, the direct relationship between RC spot counts and dose rate becomes highly nonlinear and unreliable due to variability in 3D CNN performance. In contrast to the 3D CNN training dataset, the MLP training dataset includes a broader range of challenging surface textures and high-illumination conditions, where significant false signals are present. Incorporating such conditions is essential for training the MLP to robustly identify and mitigate these false signals, thereby improving dose rate prediction accuracy in realistic settings.

For scenarios with dim lighting conditions or objects with relatively simple surface patterns, the correlation between the signal counts and dose rate demonstrates strong linearity, as illustrated in Figure \ref{fig:linear}b. This linearity arises primarily from the minimal occurrence of false positives, with an observed FPR of less than 0.03\% and recall greater than 97\%. The apparent "step-like" distribution of data points results from the discrete dose rate intervals selected during mobile phone data acquisition for training the MLP model. The theoretical dose rate detection limitation, as determined by the FPR and calculated in subsection 4.4, is approximately 10 mRem/h. A higher radiation fluence emitted from a radioactive source intensifies interactions within the CMOS image sensor, leading to an increased total number of identified signals across the entire input video sequence. Generally, radiation emitted from an examined source tends to distribute uniformly, while interactions within the CMOS sensor occur randomly, thus producing a linear relationship between dose rate and signal count, represented by the equation: Dose rate $= [1.22 \times (\text{Signal Counts}) + 3.88]\,\text{mRem/hour}$, with a coefficient of determination ($R^2$) of 0.92. The relative uncertainty associated with data points exceeding the detection limit is 22\%, with a 95\% confidence interval ranging from 1.72\% to 53.62\%, a performance superior to that of standard ion chamber survey meters.

\threesubsection{Independent external tests}

To further validate the dose rate prediction capability, two independent external tests were conducted: (i) A comparatively low-activity source (37 kBq) and (ii) A high dose rate (up to 690 mRem/h) Taylor Source. Both tests employed Cs-137, a radionuclide more likely to cause public exposure in accidental scenarios. Compared to Am-241, Cs-137 is a prevalent fission product in nuclear reactors and commonly used as a gamma radiation source in medical radiotherapy and various industrial devices. Moreover, due to its water-soluble compounds, Cs-137 exhibits higher environmental mobility, readily dispersing through air and water \cite{EPA_Cs137_Basics}.

\begin{figure}[H]
\centering
\includegraphics[width=0.9\textwidth]{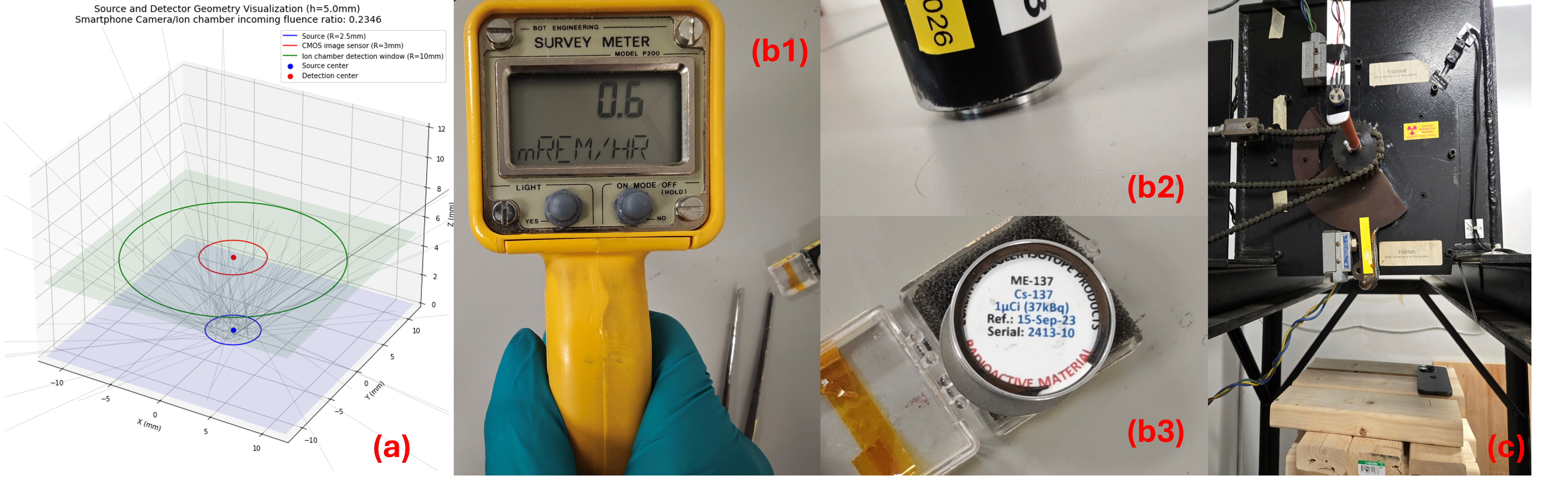}
\caption{The independent external tests setup visualization: (a), Monte Carlo simulation geometry visualization of test (i); (b1), the ion chamber dose meter reading for test (i) measurement; (b2 \& b3), the detectors and disk-shaped Cs-137   source were directly attached to each other; (c), the test (ii): the Taylor Cs-137 source measurement illustration: the black mobile phone is set below the source.}
\label{fig:external-exam}
\end{figure}

In test (i), the detectors and disk-shaped Cs-137 source were directly attached to each other, as illustrated in Figure \ref{fig:external-exam}-b2 and Figure \ref{fig:external-exam}-b3, and placed inside a cardboard box to ensure dark conditions. The measurement duration was correspondingly extended from 6 seconds (160 frames) to 50 minutes (74,900 frames), resulting in the detection of 42 signals. Under completely dark conditions, the false positive rate (FPR) of the 3D CNN was zero for the maximum achievable video duration of 1 hour. The calculated dose rate at the mobile phone camera position was 0.12 mRem/h, whereas the ion chamber dose meter reported 0.6 mRem/h, as shown in Figure \ref{fig:external-exam}-b1. By considering the effective detection areas (28 mm\textsuperscript{2} for the mobile phone camera, 314 mm\textsuperscript{2} for the ion chamber), the active source area (20 mm\textsuperscript{2}), and their approximate separation ($\sim$5 mm), a radiation fluence ratio of 0.23 between the mobile phone camera and ion chamber was derived using Monte Carlo simulation, depicted in Figure \ref{fig:external-exam}-a. Consequently, the equivalent dose rate determined by the mobile phone for the identical detection area is 0.52 mRem/h, corresponding to an accuracy of 87\%. This test is designed for possible relatively less-emergent cases, for example, to determine the radiation risk of purchased "Energy Stone" or "Healing Crystal".\\

For test (ii), the mobile phone was positioned 27 cm below and 8 cm to the side of the Taylor source window center under room-lighting conditions, as depicted in Figure \ref{fig:external-exam}-c. According to ion chamber measurements, the estimated dose rate at the mobile phone's location was 281 mRem/h. The multimodal learning model predicted a dose rate of $290\pm10$ mRem/h over eight consecutive 6-second video recordings, corresponding to an average prediction accuracy of 96.8\%. This independent validation clearly supports the reliability of the multimodal detection approach for emergency alert.

Nevertheless, additional external testing---such as outdoor experiments under strong sunlight---and further expansion of the training database are essential prior to publicly releasing a robust mobile application. Despite these remaining steps, the proposed method demonstrates substantial potential for assisting individuals in effectively avoiding radiation hazards in emergency scenarios, as well as identifying suspicious objects emitting relatively low radiation dose rates. Ultimately, this technology could facilitate timely emergency alerts for hazardous ionizing radiation exposure, thereby significantly enhancing public safety during radiological incidents.

\section{Conclusion}

This study introduces a mobile phone-based multimodal deep learning framework for real-time detection and assessment of radiation hazards, addressing a critical public safety challenge arising from the lack of widespread access to radiation detection instruments. Utilizing the intrinsic responsiveness of mobile phone CMOS image sensors to ionizing radiation, the developed approach employs a Hybrid 3D-2D Spatio-Temporal CNN capable of identifying sparse radiation-caused signals (RC spots) in raw video data, combined with brightness-based multimodal analysis through a multi-layer perceptron for dose rate prediction. This multimodal learning strategy proved crucial in mitigating false-positive detections induced by variable lighting and surface conditions. Comprehensive physics simulations using the Allpix\textsuperscript{2} framework validated the feasibility of CMOS sensors as effective radiation dosimeters, revealing a strong linear correlation between radiation fluence and sensor signal generation.

Experimental validation involving controlled radiation sources of varying activities demonstrated the proposed method's reliable performance. The model effectively detected hazardous dose rates ($>$25 mRem/h) rapidly within a brief measurement interval (six seconds). By extending the measurement period under low-activity source conditions, the system achieved an accuracy of 87\%, indicating excellent potential even in less critical scenarios ($\sim$0.6 mRem/h). 

Current limitations include a relatively small annotated dataset constrained by the sparse occurrence of radiation signals and practical radiation safety concerns during data acquisition. Future improvements necessitate the expansion of annotated databases, incorporating diverse real-world scenarios such as outdoor environments with varying illumination, and further algorithmic refinement to enhance accuracy and reduce false positives.

Ultimately, the presented mobile phone-based radiation detection method demonstrates significant promise as a practical tool to rapidly assess radiation exposure, thereby assisting individuals in avoiding harmful radiation during emergency situations. Further development and validation are essential steps toward realizing a robust and widely deployable mobile application, which could significantly enhance public situational awareness and radiation safety during radiological incidents.


\section{Experimental Section}

\subsection{Allpix\textsuperscript{2} Simulation}

Allpix\textsuperscript{2} is a parameterized simulation framework that the user defines the key modules for the simulation chain. In this study, a generic monolithic $500 \times 500$ silicon pixel detector was constructed with a pixel size of $2 \times 2\,\mathrm{\mu m}^2$ to simulate a CMOS image sensor. The senor is set to $30\,\mathrm{\mu m}$ thick and place on a support board. A simple linear electric field model is applied to the simulated sensor. The detector itself is operated under a bias voltage of 1.8\,V, while its depletion voltage is set to 1\,V. The charge carriers are propagated through the detector's material at a standard temperature of 293\,K (20$^\circ$C), a factor that directly influences their mobility and diffusion characteristics. To take account of charge sharing or crosstalk between neighboring pixels, where a single particle hit can distribute its signal across multiple pixels due to inter-pixel capacitance, a simple inter-pixel capacitance (IPC) matrix was implemented into the Capacitive Transfer module \cite{Hilbert2011_WFC3IPC}. The source was positioned 1\,cm away from the sensor and configured as a circular beam with a 0.5\,mm radius, directed along the positive Z-axis for perpendicular incidence upon the sensor. Monoenergetic 59.5\,keV gamma rays were simulated. For the intricate physics of particle interactions, the FTFP\_BERT\_LIV physics list has been selected for Geant4. In the Default Digitizer module, a default electronics noise level of 110\,electrons (e) is incorporated into the simulation and a threshold of 300\,e is set, meaning any collected pixel charge falling below this value would not be registered. All simulation results are stored in a ROOT file for data analysis.

\subsection{Equipment Setup Section}

\begin{figure}[H]
\centering
\includegraphics[width=0.9\textwidth]{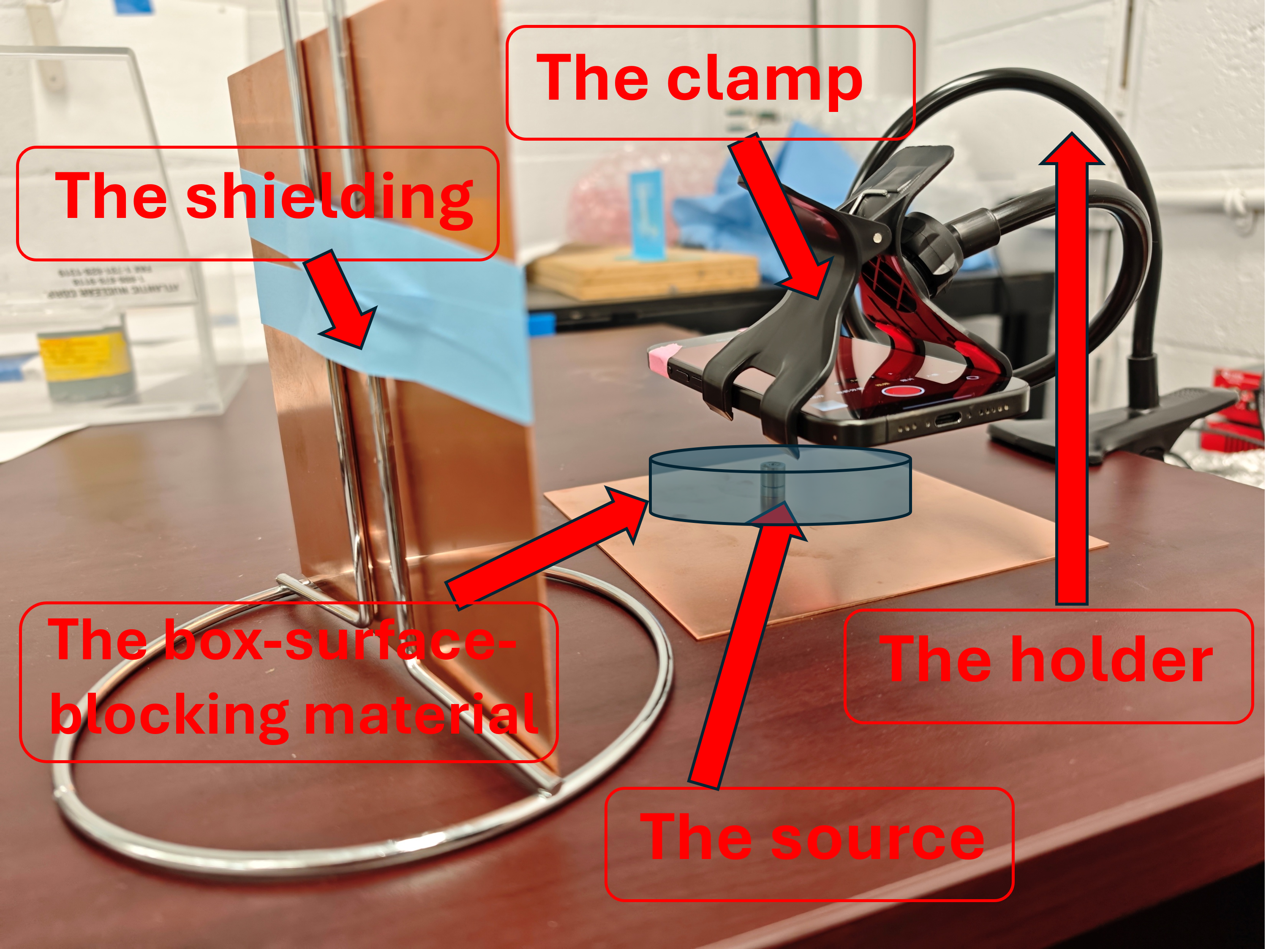}
\caption{Data measurement setup: the mobile phone is firmly fixed by the clamp, and the clamp is connected with the height-adjustable holder. The source is set right below the mobile phone camera. The box-surface-blocking material is placed between the source and the mobile phone during the data acquisition. The copper shielding is set between the source and the measurement operator to reduce the dose received by the operator's chest.}
\label{fig:setup}
\end{figure}
All data collection in this study was conducted using an iPhone 15 Pro, employing its ultra-wide camera for all video recordings. Benchmark dose rate measurements were performed with a pistol-type ion chamber survey meter (RM-P200, BOT ENGINEERING LTD). Due to handheld operation, unavoidable slight movements and the inherent sensitivity of dose rates to small variations in distance, the general uncertainty of these measurements is relatively high, approximately 30\%. An exception is the independent external test (b), where the detection window of the ion chamber was directly placed on the source surface, resulting in a more stable reading with an uncertainty of 0.1\,mRem/h, or a relative uncertainty of 17\%.\\

The box-surface-blocking method was primarily utilized for database construction and measurements. Given the strong capability of CNNs in extracting visual features, directly recording videos of radioactive sources or radiation-emitting objects might lead the trained model to incorrectly rely on distinct visual patterns rather than the sparse radiation-induced signals (RC spots). For instance, the model could potentially learn to recognize the trefoil radiation warning symbol from the positive samples instead of detecting radiation signals. To prevent this issue, various box-shaped or thin-card materials were placed between the mobile phone and the radioactive source during data acquisition for both positive and negative samples. These materials included, but were not limited to, memo paper, diverse types of cards, shipping boxes, plastic bags, disposable masks, and toilet paper. All these box-surface-blocking materials significantly reduce visible light transmission while minimally attenuating X-rays and gamma rays. Their diversity closely simulates the variety of real-world object surfaces that might be encountered in practical scenarios. Different lighting conditions were established by adjusting indoor illumination levels. Since the exact luminance value is less relevant than the brightness captured by the camera, brightness for the MLP model training was calculated using the standard formula: Brightness $= 0.299 \times Red + 0.587 \times Green + 0.114 \times Blue$ \cite{W3C_AERT_2000}.

To maximize the number of true positive RC-spot samples while maintaining radiation safety, a 555 MBq Am-241 radioactive source was employed throughout the database generation process. The mobile phone was securely mounted on a height-adjustable holder with the camera continuously recording video, while the small cylindrical source was positioned directly beneath the camera, as illustrated in Figure \ref{fig:setup}. Although the mobile phone remained firmly attached to the clamp, slight vibrations of the holder occurred upon applying gentle force, simulating realistic handheld conditions and enhancing the model's robustness against camera jitter. Each adjusted height setting corresponded to a distinct dose rate. The manual source repositioning resulted in a considerably high dose rate (6--24\,mSv/h) received by the operator. Consequently, the duration of height adjustment procedures was strictly minimized. This limitation resulted in only nine discrete dose-rate levels (including zero) being included in the MLP training dataset, accounting for the observed "step-like" distribution of data points in Figure \ref{fig:linear}b. To generate diverse databases, various box-surface-blocking materials were successively placed between the mobile phone camera and the radioactive source using 20\,cm tweezers. Each material remained briefly in position before replacement. This sequential placement and removal procedure was repeated for 15 different materials at a fixed source height to comprehensively construct the database.

\subsection{Efficient positive data selection}

\begin{figure}[H]
\centering
\includegraphics[width=0.9\textwidth]{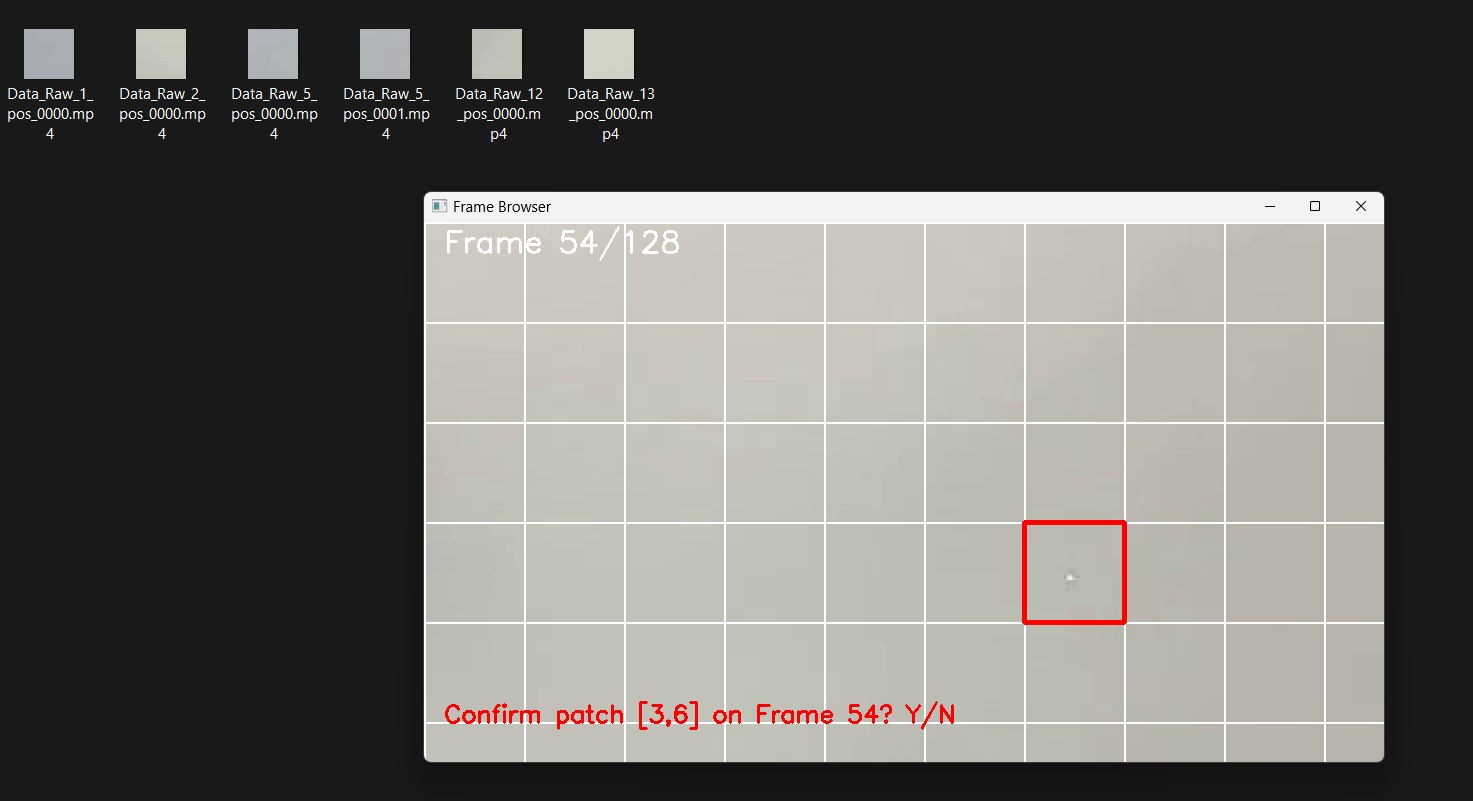}
\caption{The positive signal selection program interference window illustration on the right-below corner and the 6 automatically saved positive small video segmentations on the left-above corner.}
\label{fig:exe}
\end{figure}

After all raw video data are acquired, directly segmenting them into smaller clips of size $50\times50\times3$ channels$\times8$ frames for manual annotation is inefficient for two primary reasons: individually annotating each small video segment is excessively time-consuming, and 8-frame clips are too brief to allow effective annotation. To address this issue, an executable software tool was developed to facilitate the positive data selection process efficiently. The raw videos were initially segmented into medium-sized clips of $480\times270\times3$ channels$\times160$ frames, providing a convenient size for on-screen data annotation. As illustrated in Figure \ref{fig:exe}, this program was employed to visually select smaller clips ($50\times50\times3$ channels$\times8$ frames), indicated by red grids, containing RC spots. The selected smaller clips were automatically extracted and saved as positive and all the unselected as negative samples based on user selection.

The development of this positive data selection tool significantly reduced the time required for database construction to a feasible level. Nevertheless, due to the inherent sparsity and rarity of RC spot occurrences, database generation remained time-intensive. Specifically, annotating approximately 1000 positive small video segments ($50\times50\times3$ channels$\times8$ frames) required roughly 200 hours of manual labor by a single annotator. This substantial time commitment, coupled with radiation safety considerations, inherently constrained the size of the positive sample database. Future efforts should aim to expand the database to enhance the performance of the 3D CNN, particularly by further reducing the FPR, thereby facilitating the development of a more reliable and robust mobile phone-based radiation emergency detection application.

\subsection{Quantification and Calculation Methods}

\threesubsection{The signal–brightness correlation coefficient and the signal uniformity metric}
Seven statistical descriptors were extracted to assist MLP dose rate estimation: total signal count, mean and standard deviation of both signal and brightness, the signal–brightness correlation coefficient, and the signal uniformity metric. The \textit{signal--brightness correlation coefficient} ($\rho_{SB}$) evaluates the spatial linear relationship between the signal distribution map $S$ and the brightness map $B$ (Equation.~\ref{eq:corr}). High positive values indicate that signal intensity follows brightness patterns, which is often associated with visually induced false positives. The \textit{signal uniformity metric} ($U_S$) measures the relative dispersion of the signal distribution (Equation.~\ref{eq:uniformity}). True radiation-induced RC spots tend to be more uniformly distributed across the sensor, leading to lower $U_S$ values, whereas false signals arising from surface features or lighting variations typically yield higher $U_S$. These metrics, computed for each video segment, are integrated into the MLP modal as auxiliary inputs to enhance robustness against illumination-related artifacts.

The \textit{signal--brightness correlation coefficient} $\rho_{SB}$ quantifies the spatial linear correlation between the signal distribution map $S$ and the brightness distribution map $B$, computed as the Pearson correlation coefficient:

\begin{equation}
\rho_{SB} = \frac{\sum_{i=1}^{M} (S_i - \bar{S})(B_i - \bar{B})}{\sqrt{\sum_{i=1}^{M} (S_i - \bar{S})^2} \cdot \sqrt{\sum_{i=1}^{M} (B_i - \bar{B})^2}}
\label{eq:corr}
\end{equation}
where $S_i$ and $B_i$ are the pixel values at position $i$, $\bar{S}$ and $\bar{B}$ are their spatial means, and $M$ is the total pixel count.

The \textit{signal uniformity metric} $U_S$ characterizes the dispersion of the signal map:

\begin{equation}
U_S = \frac{\sigma_S}{|\bar{S}| + \epsilon}
\label{eq:uniformity}
\end{equation}
where $\sigma_S$ and $\bar{S}$ denote the standard deviation and mean of $S$, respectively, and $\epsilon$ is a small constant to avoid division by zero.

\threesubsection{Critical Level and Detection Limit }

Following Lloyd A. Currie's work \cite{Currie1968,Currie2004} on limits of detection, we define the net counts as $N = C_G - C_{\mathrm{FP}}$, where $C_G$ is the gross count and $C_{\mathrm{FP}}$ is the expected false-positive (FP) count. Replacing background counts by FP counts reflects our objective of discriminating true RC-spot signals from visually induced artifacts. Under Poisson statistics and equal timing, $\mathrm{var}(N)=\mathrm{var}(C_G)+\mathrm{var}(C_{\mathrm{FP}})=(C_N)+2C_{\mathrm{FP}}$, yielding the standard deviation:
\begin{equation}
\sigma_0=\sqrt{2\,C_{\mathrm{FP}}}. 
\end{equation}
The \textit{critical level} for a certain one-sided mistake risk $\alpha$ is then
\begin{equation}
L_C = k_\alpha \,\sigma_0 = k_\alpha \sqrt{2\,C_{\mathrm{FP}}}.  \label{eq:LC-FP}
\end{equation}
The risk of mistake $\alpha$ corresponds to a $k_\alpha$ value, for example, $k_\alpha=1.282$ corresponds to $\alpha=0.1$ for 90\% one-side degree of confidence.
This adapts the standard form $L_C=k_\alpha\sigma_0$ with $\sigma_0$ evaluated from the FP-only distribution rather than background-only measurements.

The \textit{detection limit} $L_D$ is defined to control the miss probability $\beta$ when the true net count equals $L_D$. For equal one-sided risks ($k_\alpha=k_\beta=k$), the derivation of Currie's previous work \cite{Currie1968} gives the following.
\begin{equation}
L_D = k^2 + 2k\,\sigma_0 \;\;\approx\;\;  2k\,\sqrt{2\,C_{\mathrm{FP}}}, \label{eq:LD-FP}
\end{equation}

In practice, $C_{\mathrm{FP}}$ per video is estimated from the 3D-CNN false positive rate ($\mathrm{FPR}$) and the number of scanned segments $N_{\mathrm{seg}}$: 
\begin{equation}
C_{\mathrm{FP}} \approx \mathrm{FPR}\times N_{\mathrm{seg}}.
\end{equation}
For hazard screening, counts thresholds $L_C$ and $L_D$ can be converted to dose-rate thresholds via the empirically linear mapping between RC-spot counts and dose rate in low-FP regimes.

\medskip
\textbf{Supporting Information} \par 
Supporting Information is available from the Wiley Online Library or from the author.

\medskip
\textbf{Acknowledgements} \par 
The authors would like to thank McMaster University Health Physics, especially Jennifer Clarke, for providing professional radiation protection supervision and guidance, and an independent researcher, Ding Chaoyang, for guidance on raw data extraction of CMOS image sensors. The authors also would like to thank Liu Wei of the Radiation Sciences Graduate Program of McMaster University for helping with the measurement of the Taylor gamma source and Prof. Soo Hyun Byun of the Radiation Sciences Graduate Program of McMaster University for helping with the draft review and edition. 

\medskip
\textbf{Funding} \par
The authors would like to thank Mitacs and Laurentis Energy Partners for financially supporting this project.

\medskip
\textbf{Conflict of Interest} \par
The authors declare no conflict of interest.

\medskip
\textbf{Data Availability Statement} \par
The data that support the findings of this study are available from the cor
responding author upon reasonable request.

\medskip

%

\bibliographystyle{unsrtnat} 
\bibliography{references}


\end{document}